\documentclass[twocolumn]{aastex63}
\usepackage{ulem,color}
\usepackage{amsmath,amssymb,bm}
\usepackage{etoolbox}
\usepackage{booktabs}
\usepackage{natbib}
\def \Ohydro {\Omega_{\rm hydro}}
\def \Gn   {\Gamma_{\rm 0}}

\def \hchi  {\hat{\chi}}
\def \KK  {\bm{K}}
\def \cnot  {c_{\rm 0}}
\def \rhon {\rho_{\rm 0}}
\def \Pn   {P_{\rm 0}}
\def \Bnot {B_{\rm 0}}

\def \Bn  {B_{\rm 0}}

\def \rhon {\rho_{\rm 0}}

\def \del2z {\partial^{2}_{z}}

\def \vA   {v_{\rm A}}

\def \A {{\bm A}}

\def \f {{\bm f}}

\def \uu {{\bm u}}
\def \bb {{\bm b}}
\def \UU {{\bm U}}

\def \B {{\bm B}}

\def \JJ {{\bm J}}
\def \mun {\mu_{{\rm 0}}}

\def \k {{\bm k}}
\def \x {{\bm x}}

\def \curl {{\bm \nabla} \times}
\def \dive {{\bm \nabla}\cdot}

\def \delt {\partial_t}

\def \t  {\theta}
\def \p  {\phi}

\newcommand{\eq}[1]{(\ref{#1})}

{}

\def \MaA  {\mbox{M}_{\rm A}}

\def \BB {\bm B}
\def \B {\bm B}

\def \A {\bm A}

\def \curl {{\bm \nabla}\times}

\def \BB {\bm B}

 \def \a {{\alpha_0}}

\renewcommand{\fig}[1]{Fig.~\ref{#1}}

\newcommand{\epsn}{{\bold{\color{black}\epsilon}}}

\def\drawing #1 #2 #3 {
\begin{center}
\setlength{\unitlength}{1mm}
\begin{picture}(#1,#2)(0,0)
\put(0,0){\framebox(#1,#2){#3}}
\end{picture}
\end{center} }

\usepackage[normalem]{ulem}

\usepackage[vskip=1em,font=itshape,leftmargin=1em,rightmargin=1em]{quoting}

\usepackage{bm}

\shorttitle{Anisotropic Magnetized Asteroseismic Waves}
\shortauthors{Tripathi and Mitra}

\begin{document}
\title{Anisotropic Magnetized Asteroseismic Waves}
\author[0000-0002-4723-2170]{B.~Tripathi}
\affiliation{Department of Physics, University of Wisconsin--Madison, Madison, Wisconsin 53706, USA}
\email{btripathi@wisc.edu}

\author[0000-0003-4861-8152]{Dhrubaditya Mitra}
\affiliation{Nordita, KTH Royal Institute of Technology and Stockholm
  University, Hannes Alfv\'ens v\"ag 12, 114 19 Stockholm, Sweden}
\email{dhruba.mitra@gmail.com}

\begin{abstract}
  We solve for waves in a polytropic, stratified plasmas with a spatially
  varying background magnetic field that points along a horizontal
  $x$-direction, and with gravity that is directed along the vertical
  $z$-direction.
  Force balance determines the magnitude of the background magnetic field,
  $\Bnot^2 \sim z^{n+1}$, where $n$ is the polytropic index.
  Using numerical and asymptotic methods, we deduce
  an explicit dispersion relation for fast pressure-driven waves:
  $\Omega^2  \sim  K\left(2m+n\right) \left[1 + (1/\MaA)^2 (4-2\gamma+\cos^2\theta-3\cos^4\theta)/4 \right]$.
  Here, $\Omega$ is the frequency, $K$ the wavenumber, $\theta$ the angle the
  wave-vector makes with the background magnetic field, $\MaA$ the
  Alfv\'enic Mach number, and $m$ an integer representing the eigenstate.
  We discuss roles of such an explicit formula in asteroseismology.
\end{abstract}


\section{Introduction} 

The strengths of magnetic fields buried below the surface of stars are not
known, though they are vital for improved understanding of the magnetic
behaviour of stars.
This challenge has impeded progress in understanding
stellar magnetism and the evolution of magnetized stellar interiors.
To estimate the magnetic field strength, linear asteroseismology is a promising technique~\citep{aerts2010}.
The key idea is to calculate the dispersion relations of surface waves
taking into account the presence of subsurface magnetic fields.
Then inversion techniques are used, leveraging the observed power spectrum of oscillations,
to infer the subsurface magnetic field.
Thus, simple analytical dispersion relations are insightful, but obtaining such
relations remains a challenge.
This challenge also impacts the development of nonlinear
asteroseismology~\citep{guo2020, beeck2021, beeck2023},
which requires linear dispersion relations to evaluate mode resonances.
Consequently, the progress in magnetoseismology has been slow.

Observational studies report travel-time perturbations of acoustic waves to
be a critical signature of strong magnetic fields in the stellar
interior~\citep{schunker2005, ilonidis2011}.
Numerical simulations of asteroseismic waves also suggest the possibility of
detecting subsurface fields, before they emerge on the
surface~\citep{singh2014, singh2015pfmodes, singh2016high, singh2020,
  das2020sensitivity, das2022recipe}.
To bolster such findings, a thorough understanding of the impact of magnetic
fields on asteroseismic waves is essential~\citep{nye1976,adam1977occurrence,
  thomas1983magneto, campos1983three, cally2007, campos2015combined,
  tripathi2022a}.

Waves in an unmagnetized polytropic atmosphere were exactly solved analytically
by \cite{lamb1911} who derived the relation
\begin{equation}
  \label{eq:lambfull}
  \frac{\Omega^2}{2K} - \frac{(n+1)(n + 1 - \gamma n )K}{2\gamma^2 \Omega^2} =
  m+\frac{n}{2},
\end{equation}
where $\Omega$ is the frequency,  $K$ the wavenumber,
$n$ the  polytropic index,
$\gamma$ the adiabatic index, and $m$ the eigenstate index, with
$m=0,1,2,...\,$. 
This advancement led to a series of newer and significant understanding of
hydrodynamic waves.
Under fast-wave approximation, $\Omega^2/K\gg 1$, the leading-order dispersion relation becomes
\begin{equation}
  \label{eq:lambasymp}
  \Omega^2  \sim K \left(2m+n\right)\/.
\end{equation}

A similar closed-form analytical expression for waves in a magnetized
polytrope, as of yet, is unknown.
Earlier attempts \citep[e.g.,][]{gough1990, spruit1992, cally1993,
    bogdan1997, cally1997simulation} have ended with
  deriving, analytically, approximate integral expression
  in certain perturbative limits and then solving them numerically. 
  Hence, they leave out the critical step of obtaining a straightforward
  analytical understanding of the effect of magnetic fields on linear
  asteroseismic waves.
  For example, \cite{spruit1992} consider a plane-parallel polytropic
  atmosphere threaded by a vertical magnetic field that is uniform throughout
  the whole domain.
  They perturbatively derive integrals for changes in wave frequency and then
  solve the integrals numerically.
  \cite{cally1993} pursue the model of \cite{spruit1992}, but with a different
  perturbative approach.
  The effect of horizontal fields on waves is studied by
  several authors~\citep[e.g.,][]{nye1976,adam1977occurrence,
  thomas1983magneto, campos1983three, cally2007, campos2015combined,
  tripathi2022a},
  but in  simple and unrealistic isothermal atmosphere, which has very
  different properties of waves than in a more realistic polytropic atmosphere.
  Notably, \cite{gough1990} treat a global problem (in spherical coordinates);
  there, the computation of eigenfrequencies requires evaluation
  of integral expressions using numerical methods [e.g., Eqs.~(4.11)--(4.13)
    in \cite{gough1990}].
  A closed-form analytical formula is currently unavailable. 
The limitations of purely numerical approach
and lack of a closed-form expression were succinctly expressed
by \cite{bogdan1997}:
\begin{quote}
``\textit{Ideally, we would wish to proceed by writing down an equation analogous
to Lamb’s formula  for the magnetized polytrope. Unfortunately, this
approach is not feasible and for the most part one must instead be content
with a numerically derived visual comparison of how the allowed oscillation
frequencies depend upon the choice of the horizontal wavenumber $k$.}''
\end{quote}

Analytic dispersion relations are also essential for developing
wave turbulence theory in the presence of both gravity and magnetic fields.
In wave turbulence theory, calculations of mode resonances require simple,
explicit dispersion relations that accurately capture the magnetic effect on
observed linear waves.
We note that the Lamb's dispersion relation~\eq{eq:lambasymp},
$\Omega\sim\sqrt{K}$, is similar to that of surface gravity waves in
oceans~\citep{hasselmann1962}.
However, there is a critical difference:  The surface gravity waves do not
couple via three-wave resonance, thus requiring a weaker
four-wave coupling~\citep{nazarenko2016}.
The Lamb waves, on the other hand, can couple via three-wave resonance,
because there are infinitely many such waves (eigenstates) at a given wavenumber, unlike only one pair of surface gravity waves at a given wavenumber.
Thus, the infinitely many Lamb waves at a given wavenumber have distinct wave
frequencies, which allow the sum of three frequencies at three wavenumbers to
become null.
While a wave turbulence theory for surface gravity waves has been well
established, it is yet to be developed for the asteroseismic waves,
whose dispersion relation in fully analytic form is a basic requirement for
such a theory.
The value that an analytic dispersion relation offers in resonant-coupling
theory cannot be overstated when magnetic fields make the wave dispersion
relation anisotropic and complicated.
Motivated by these reasons, we seek here an accurate and simple formulae for
the effect of magnetic fields on the Lamb waves.

Introducing magnetic fields, aligned orthogonal to a vertical gravity
in Fig.~\ref{fig:f1}, we find, as did \cite{bogdan1997}, that the linearized
magnetohydrodynamic (MHD) equations are too cumbersome to obtain
a closed-form expression for the dispersion relation, even with the
fast-wave approximation,  $\Omega^2/K\gg 1$.
Here, we overcome this difficulty by using both numerical simulations and
extensive use of Mathematica, followed by a variant of the
Jeffereys--Wentzel--Kramers--Brillouin (JWKB) approximation
we devise to deduce
\begin{equation}\label{eq:principalresult}
    \Omega^2  \sim  K\left(2m+n\right)\left[1 + \frac{\epsn^2 (4-2\gamma+\cos^2\theta-3\cos^4\theta)}{4} \right],
\end{equation}
in the limit $\Omega^2/K\gg 1$, with $m=0,1,2,...\,$.
The parameter $\epsn$ is the inverse of the Alfv\'enic
Mach number, and $\theta$ is the angle the wave vector makes with the
background magnetic field.
Equation~\eqref{eq:principalresult} is the principal result of this paper.

This paper is organized in the following way. In \S~\ref{sec:setup}, we
describe our model and present the linearized compressible MHD equations.
Such equations are then numerically solved in \S~\ref{sec:comparenumanalytical}.
To obtain analytical understanding of the numerical results, the linearized
equations are reduced to a wave equation in \S~\ref{sec:nondim}.
The normal-form wave equation is then perturbatively solved using a variant of
the JWKB theory we construct; analytical understanding is gained
in \S~\ref{sec:pertsol}.
With astrophysical implications and roles of our results, we conclude
in \S~\ref{sec:conclusion}.

\section{System setup and linearized perturbation equations}
\label{sec:setup}
\begin{figure}[t!]
    \centering
    \includegraphics[width=0.4\textwidth]{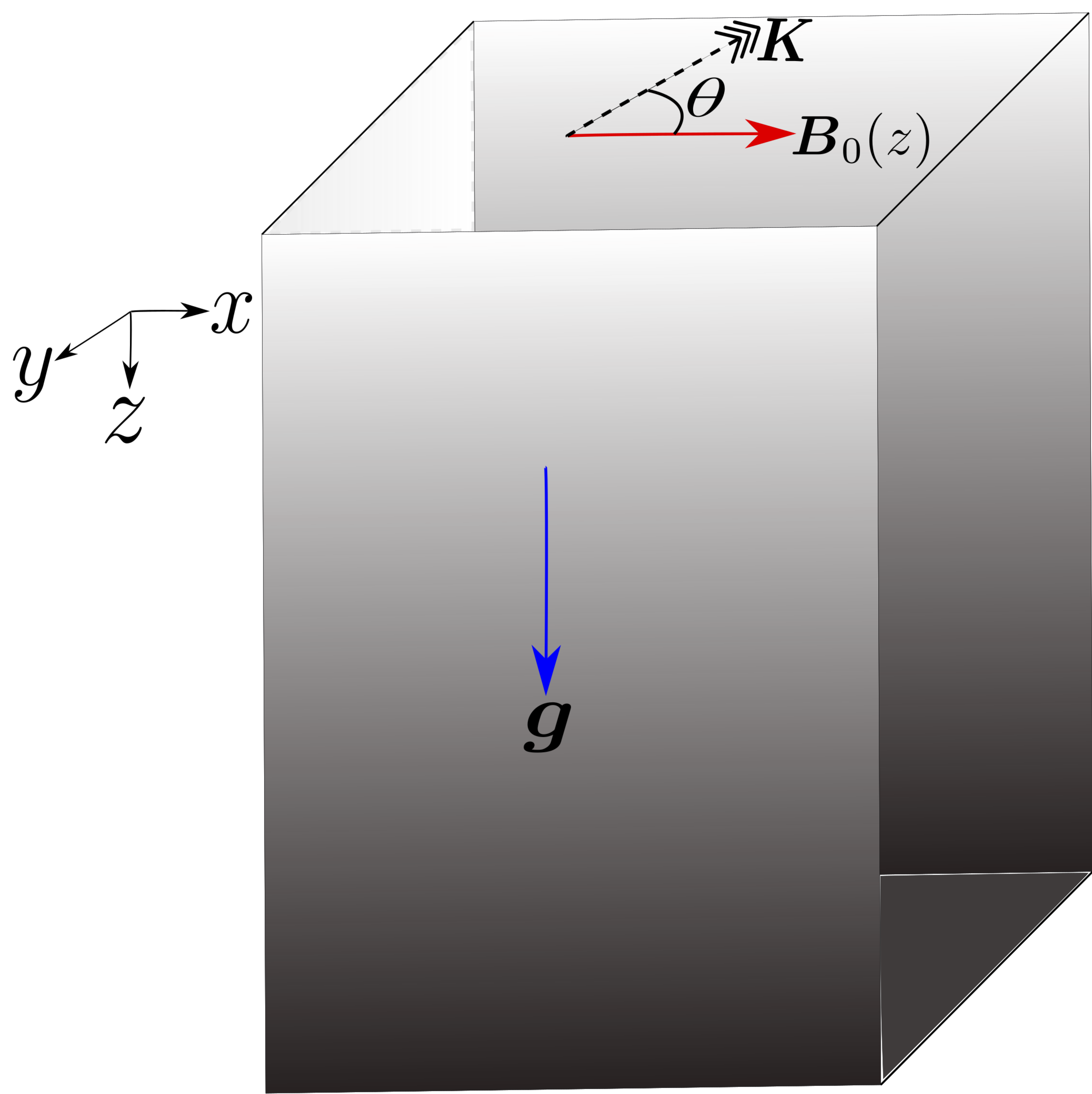}
    \caption{An inhomogeneous magnetic field $\bm{B}_0(z)$, oriented orthogonal
      to a constant vertical gravity $\bm{g}$, is considered where the wave is
      allowed to propagate in an arbitrary direction, shown with $\bm{K}$,
      making an angle $\theta$ with $\bm{B}_0$. The gradient in the colormap
      of the box schematically represents increasing functions with depth
      $z$ of the magnetic field, fluid density, pressure, and sound speed.}
    \label{fig:f1}
\end{figure}
To study waves in a magnetized, stratified medium, we consider the ideal three-dimensional compressible magnetohydrodynamic (MHD) equations \citep{chandra1961}
\begin{subequations}
  \begin{align}
    \delt \rho + \dive \left(\rho \UU \right) &= 0  \/,
    \label{eq:cont} \\
    \rho \left[\delt \UU + (\UU \cdot \bm{\nabla})\UU\right]  &= -\bm{\nabla} P + \rho \bm{g} + \bm{J} \times \bm{B} \/,
    \label{eq:mom}\\
    \delt \BB &= \curl\left( \UU\times\BB \right),\label{eq:induction}\\
    \frac{Dp}{Dt} &= c^2 \frac{D\rho}{Dt}, \label{eq:eos}
  \end{align}
  \label{eq:MHD}
\end{subequations}
where $\rho$, $\UU$, $P$, and $\BB$ represent the density, the velocity, the pressure, and the magnetic field, respectively. The current density is $\JJ=\curl\BB/\mun$, where $\mun$ represents the magnetic permeability of the vacuum. The magnetic field is additionally constrained to be divergence-less
\begin{equation}
      \dive \BB = 0.
\end{equation}
We consider a local Cartesian domain where the equilibrium density $\rhon$ and pressure $\Pn$
satisfy the polytropic relation
\begin{equation}\label{eq:poly}
    \Pn \sim \rhon^{1+1/n},
\end{equation}
where $n$ is the polytropic index of the gaseous atmosphere.
In an unmagnetized atmosphere, the force balance gives
$\partial_z \Pn(z) = \rhon(z) g$, where
$\bm{g} = g \hat{\bm{e}}_z$,
with $g$ as the constant acceleration due to gravity along the vertical $z$-axis, as depicted in Fig.~\ref{fig:f1}.
The force balance requires
\begin{equation}
  \label{eq:background_p_rho12} 
\rhon(z) \sim z^n; \quad \Pn(z) \sim z^{n+1}.
\end{equation}

\subsection{Background magnetic field and sound speed}
We introduce an inhomogeneous background magnetic field,
$\BB=\Bnot(z) \hat{\bm{e}}_x$.
The force-balance relation with the Lorentz force,
$(\curl\BB/\mun) \times \BB$, then becomes
\begin{equation}
  \label{eq:forcebalance}
    \partial_z \Pn = \rhon g- \frac{1}{2\mun}\partial_z \Bnot^2.
\end{equation}
The solution 
\begin{equation}
  \label{eq:background_Bfinal}
    B_0^2(z) \sim z^{n+1}\/
\end{equation}
also satisfies Eq.~\eq{eq:background_p_rho12}.
The plasma $\beta$ is then a constant throughout the domain
\begin{equation}
    \beta \equiv \frac{P_\mathrm{gas}}{P_{\mathrm{magnetic}}} = \frac{2\mun \Pn}{\Bnot^2}.
\end{equation}
The sound speed ($c$) may now be deduced from Eq.~\eqref{eq:forcebalance}, using
$\Pn= c^2 \rhon/\gamma$,
where $\gamma$ is the adiabatic index of the gas
\begin{equation}\label{eq:soundspeedvar}
    \partial_z c^2 =- \frac{n c^2}{z} + \frac{\gamma g}{ 1 + \beta^{-1}}.
\end{equation}
The solution to Eq.~\eqref{eq:soundspeedvar} is
$c^2 \mathrm{=} \cnot^2 z$, where $\cnot^2$ is a constant given by
\begin{equation}
    \cnot^2 = \frac{\gamma g }{(n+1) (1+\beta^{-1})}\/.
\end{equation}
It is useful to express $\beta^{-1}$ in terms of the Alfv\'enic~Mach number $\MaA$,
which is the ratio of the the sound speed to the Alfv\'en speed,
\begin{equation}\label{eq:betadefn}
  \beta^{-1} = \frac{\Bnot^2}{\mu_0 \rho_0}  \cdot \frac{\rho_0}{ \gamma \Pn} \cdot \frac{\gamma}{2}
  = \frac{\gamma}{2\MaA^2}\/,
\end{equation}
using which the sound speed becomes
\begin{equation} \label{eq:soundspeed1}
    c^2 = \frac{\gamma g z}{(n+1)(1 + \gamma\MaA^{-2}/2)}.
\end{equation}

\subsection{Linearized perturbation equations}
We now linearize the MHD equations around the background profiles $[\rhon, \bm{0}, \Bnot, \Pn]$,
introduced in Eqs.~\eqref{eq:background_p_rho12} and \eqref{eq:background_Bfinal}.
Such a linearization yields evolution equations for infinitismal perturbations
$[\widetilde{\rho}, \widetilde{\uu}, \widetilde{\bb}, \widetilde{p}]$ as
\begin{subequations}
\begin{align} \label{eq:lina}
    \partial_t \widetilde{\rho} &= - (\widetilde{\bm{u}} \cdot \bm{\nabla}) \rhon - \rhon (\bm{\nabla} \cdot \widetilde{\bm{u}})\/,\\
    \partial_t \widetilde{u}_x &= - \frac{\partial_x \widetilde{p}}{\rho_0} + \frac{\widetilde{b}_z \partial_z \Bn}{\mun \rhon}\/,\\
    \partial_t \widetilde{u}_y &= - \frac{\partial_y \widetilde{p}}{\rho_0} + \frac{\Bn(\partial_x \widetilde{b}_y - \partial_y \widetilde{b}_x )}{\mun \rhon}\/,\\
    \partial_t \widetilde{u}_z &= - \frac{\partial_z \widetilde{p}}{\rhon} + \frac{\widetilde{\rho } g}{\rhon} + \frac{\Bn(\partial_x \widetilde{b}_z - \partial_z \widetilde{b}_x ) - \widetilde{b}_x \partial_z \Bn }{\mun \rhon}\/,\\
    \partial_t \widetilde{b}_x &= - \Bn (\partial_y \widetilde{u}_y + \partial_z \widetilde{u}_z ) - \widetilde{u}_z \partial_z \Bn\/,\\
    \partial_t \widetilde{b}_y &= - \Bn \partial_x \widetilde{u}_y\/, \\
    \partial_t \widetilde{b}_z &= - \Bn \partial_x \widetilde{u}_z\/, \\ \label{eq:linh}
    \partial_t \widetilde{p} &= - (\widetilde{\bm{u}} \cdot \bm{\nabla}) \Pn - c^2 \rho_0 (\bm{\nabla} \cdot \widetilde{\bm{u}})\/.
\end{align}
\end{subequations}
We shall use 
\begin{equation}
    \widetilde{\chi}=\bm{\nabla} \cdot \widetilde{\uu} 
\end{equation} 
in the rest of the article, where helpful.

Equations~\eqref{eq:lina}--\eqref{eq:linh} can be simplified to derive a set of fewer (closed) equations
\begin{subequations}
\begin{align} \label{eq:linux}
\partial_t^2 \widetilde{u}_x &= c^2 \partial_x \widetilde{\chi} + g \partial_x \widetilde{u}_z\/,\\ \label{eq:linuy}
\partial_t^2 \widetilde{u}_y &= c^2 \partial_y \widetilde{\chi} + g \partial_y \widetilde{u}_z +
\frac{c^2}{\MaA^2} \left(\partial_{xx} \widetilde{u}_y + \partial_{y} \widetilde{\chi} - \partial_{xy} \widetilde{u}_x \right)\/,\\\label{eq:linuz}
\partial_t^2 \widetilde{u}_z &= c^2 \partial_z \widetilde{\chi} - g \partial_x \widetilde{u}_x \left[ 1 + \frac{\gamma }{\MaA^2 (1+\gamma \MaA^{-2}/2)} \right] - g \partial_y \widetilde{u}_y \nonumber \\
&\hspace{0.2cm}+\frac{\gamma (1+ \MaA^{-2}) g \widetilde{\chi}}{(1+\gamma \MaA^{-2}/2)}   + \frac{c^2}{\MaA^2} \left(\partial_{xx} \widetilde{u}_z + \partial_{z} \widetilde{\chi} - \partial_{xz} \widetilde{u}_x \right)\/.
\end{align}
\end{subequations}
We note that the appearance of $\MaA^2$ in Eq.~\eqref{eq:linuz} in certain terms may seem non-trivial at first
sight; however, upon inspection, we understand them as terms emerging from the effect of the Lorentz force on the
background states, via terms like $\partial_z c^2(z)$ and $(\partial_z \Pn)/\rho_n$, while processing
Eqs.~\eqref{eq:lina}--\eqref{eq:linh}.
Admittedly, Eqs.~\eqref{eq:linux}--\eqref{eq:linuz} are somewhat challenging to proceed with clarity.
Thus, we now non-dimensionalize the equations to make them reasonably transparent.

\subsection{Non-dimensionalized linear equations}
We define $L$ as the characteristic length scale over which the sound speed varies (and, for that matter, pressure, density, and temperature also vary) appreciably. Then we find that the characteristic sound speed in an unmagnetized polytrope is, using Eq.~\eqref{eq:soundspeed1}, $c_L=\sqrt{\gamma g L/(n+1)}$. Using $L$ and $L/c_L$ as the dimensional length and time units, we non-dimensionalize all variables henceforth, starting with the sound speed
\begin{equation}\label{eq:C2defn}
    C^2 = \frac{c^2}{c_L^2}  = \frac{Z}{1 + \gamma\MaA^{-2}/2},
\end{equation}
where the lowercase dimensional variables ($c$ and $z$) are cast as uppercase non-dimensional
variables ($C$ and $Z$).
Thus we replace all the dimensional variables in Eqs.~\eqref{eq:linux}--\eqref{eq:linuz} using
\begin{subequations}
\begin{align}
    (x,y,z) &= (LX, LY, LZ),\\
    c^2 &= c_L^2 \frac{ Z}{1 + \gamma\MaA^{-2}/2}, \\
    \partial_t &\equiv \frac{c_L}{L} \partial_T,
\end{align}
\end{subequations}
where the uppercase characters $X,Y,$ and $T$ represent non-dimensional variables.

We analyze perturbations by Fourier-transforming in the $(x,y)$-plane, viz.,
\begin{equation}
  \widetilde{u}_x(Z) = \int dK_XdK_Yd\Omega\, \hat{u}_x \exp\left[ i (K_X X + K_Y Y + \Omega T)\right],
\end{equation}
where the uppercase characters represent non-dimensional quantities, e.g.,
$\KK \equiv (K_X, K_Y)$ is the non-dimensional wavevector in the $(x,y)$-plane, and
$\Omega$ is the nondimensional frequency.
Fourier analyzing Eqs.~\eqref{eq:linux}--\eqref{eq:linuz} and representing
$\MaA^{-1}$ by $\epsn$ henceforth, we write
\begin{subequations}
\begin{align} \label{eq:linux2}
&\left[\frac{-\Omega^2(1+\gamma \epsn^2/2)}{i K_X Z}\right]\hat{u}_x + \left[\frac{-(n+1)(1+\gamma \epsn^2/2)}{\gamma Z} \right] \hat{u}_z = \hat{\chi}\/,\\ \label{eq:linuy2}
&\left[-i \epsn^2 K_X \right] \hat{u}_x + \left[\frac{-\Omega^2(1+\gamma \epsn^2/2)}{i K_Y Z} + \frac{\epsn^2 K_X^2}{i K_Y}\right]\hat{u}_y\nonumber\\
&\hspace{2cm}+ \left[\frac{-(n+1)(1+\gamma \epsn^2/2)}{\gamma Z} \right] \hat{u}_z = \hat{\chi} (1+\epsn^2)\/,\\ \label{eq:linuz2}
&\left[ i K_X\left\{ \frac{1 + \gamma \epsn^2/2}{\gamma} + \epsn^2  \left(1+ \frac{Z \partial_Z}{n+1}  \right) \right\} \right] \hat{u}_x \nonumber\\
&+ \left[\frac{i K_Y  (1 + \gamma \epsn^2/2)}{\gamma}\right]\hat{u}_y + \left[ \frac{-\Omega^2(1+\gamma \epsn^2/2) + \epsn^2 Z K_X^2 }{n+1}\right] \hat{u}_z\nonumber\\
&\hspace{3.5cm}= \left(\hat{\chi} + \frac{Z \partial_Z \hat{\chi} }{n+1} \right) (1+\epsn^2)\/.
\end{align}
\label{eq:lin}
\end{subequations}
Equations~\eqref{eq:linux2}--\eqref{eq:linuz2} are general, without any
  approximation, and host all three families of MHD waves: the fast and slow
  magnetoacoustic waves, and the Alfv\'en waves. 
  The Alfv\'en wave dispersion relation is reproduced by substituting $K_Y=0$
  and $K_X=K$ in Eq.~\eqref{eq:linuy2}, which renders it to: 
  either $\hat{u}_y=0$ (the trivial solution), or
  $\Omega^2=\epsilon^2 K^2 Z/(1+ \gamma \epsilon^2)$. 
  This is indeed the dispersion relation of the Alfv\'en waves, which is
  confirmed by writing $\Omega^2 = K^2 \vA^2$, where
  $\vA^2=\Bnot^2/(\mun \rhon)$ is the squared Alfv\'en speed. 
  This squared Alfv\'en speed is determined from Eq.~\eqref{eq:betadefn} to be
  $\gamma \Pn/(\rhon \MaA^2)$.
  Recognizing that $\gamma \Pn/\rhon$ is the squared sound speed, given in
  Eq.~\eqref{eq:C2defn} [derived from Eq.~\eqref{eq:soundspeedvar}],
  the squared Alfv\'en speed is 
  $\vA^2=\epsilon^2 Z/(1+ \gamma \epsilon^2)$.

\section{Exact numerical solution}
\label{sec:comparenumanalytical}
  We obtain fully converged numerical solution to Eqs.~\eqref{eq:linux2}--\eqref{eq:linuz2} by employing the spectral ``Dedalus" framework~\citep{burns2020}.
  Referring to the Dedalus methods paper~\citep{burns2020} for more details,
  we briefly outline the numerical procedures employed in Dedalus for eigenvalue problems.
  At each horizontal wavenumber $(K_X, K_Y)$, the state variables---the three components of velocity---are expanded in the
  Chebyshev polynomials along the inhomogeneous $z$-axis. Because of the background inhomogeneity, different
  Chebyshev coefficients couple, creating a dense linear operator. Sparsification is provided by a change of basis
  from the Chebyshev polynomials of the first kind to those of the second kind.
  To impose boundary conditions and keep the matrix sparse, Dirichlet preconditioning is applied.
  Efficient solution of the resulting matrices is then found by passing the matrices
  $\mathcal{L}$ and $\mathcal{M}$ in the eigenvalue ($\sigma$) problem,
  $\mathcal{L} \mathcal{X} = \sigma \mathcal{M} \mathcal{X}$, to the ``scipy" linear algebra packages.
For a given spectral resolution along the inhomogeneous $z$-axis,
    we solve for all the eigenvalues of the matrices.
    Such a non-targeted, general solution produces a complete eigenspectrum and
    eigenmodes, corresponding to all families of linear MHD waves:
    the (fast) pressure-driven, (slow) gravity-driven modes, and the
    Alfv\'en modes.
    We focus on the high-frequency modes to assess the effect of magnetic
    fields on such fast pressure-driven waves.
\begin{figure}
    \centering
    \includegraphics[width=0.9\columnwidth]{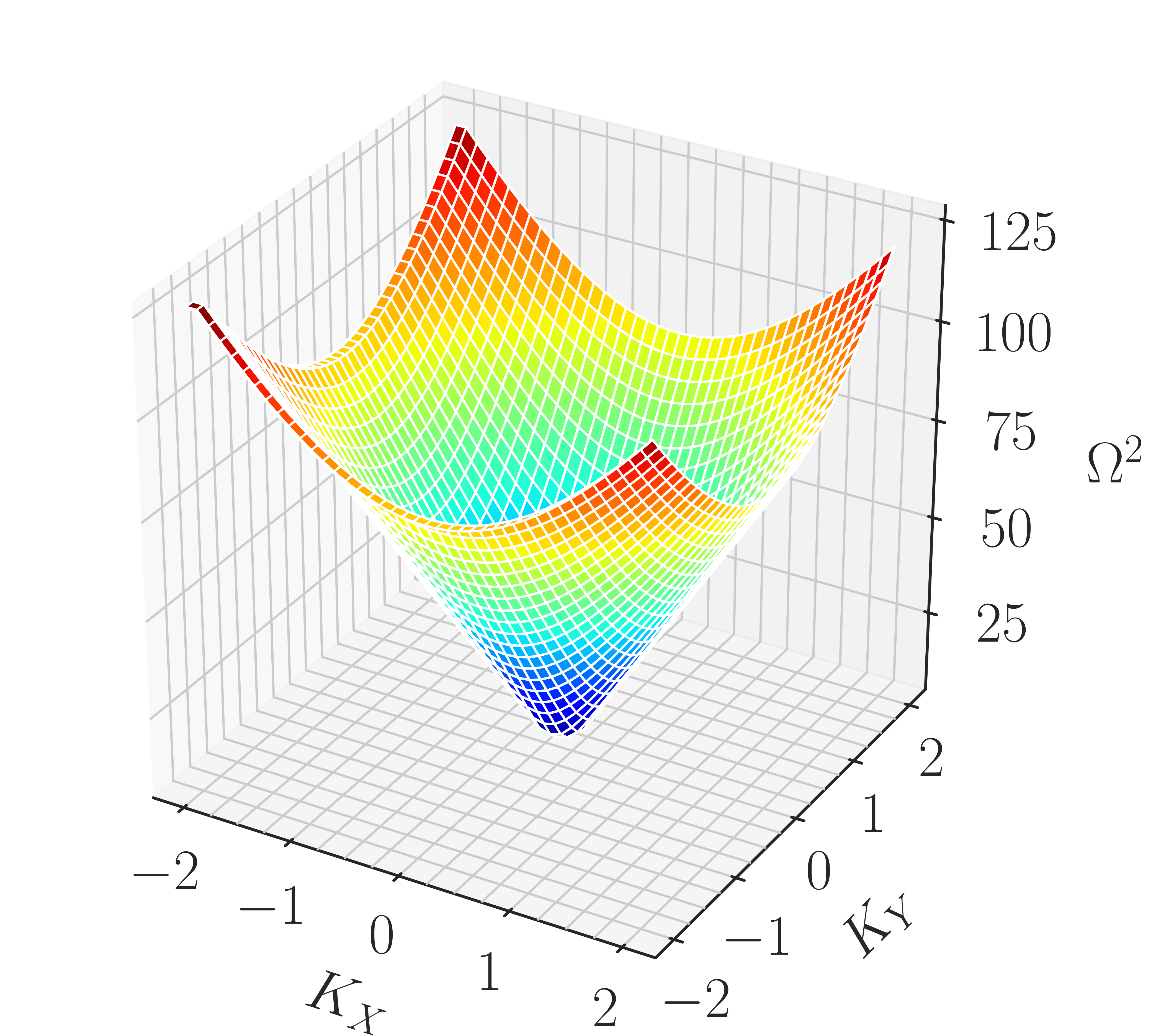}
    \caption{$\Omega^2(K_X,K_Y)$ for hydrodynamic and magnetized polytropes. The two plotted surfaces are visually indifferentiable because the difference ($\Delta\Omega^2$) between them is much smaller than $\Omega^2$; see Fig.~\ref{fig:f3}. The parameters used are $\epsilon=0.1$, $m=20$, $n=2.5$, and $\gamma=5/3$. For variations in these parameters, the surface plot of $\Omega^2(K_X,K_Y)$ remains the same qualitatively.}
    \label{fig:f2}
\end{figure}

For the boundary condition, at the lower boundary $z=L_z$, we require $\widetilde{u}_z=0$.
At the upper boundary, $z=0$, where the atmosphere ceases, we enforce, following \cite{lamb1911},
\begin{equation}
\begin{aligned}[b]
    \frac{Dp}{Dt} &= 0,
\end{aligned}
\end{equation}
which implies
\begin{equation}
    c^2 \frac{D\rho}{Dt} = -c^2 \rhon \widetilde{\chi}  = 0 = z^{n+1} \widetilde{\chi}.
\end{equation}

\begin{figure}
    \centering
    \includegraphics[width=0.9\columnwidth]{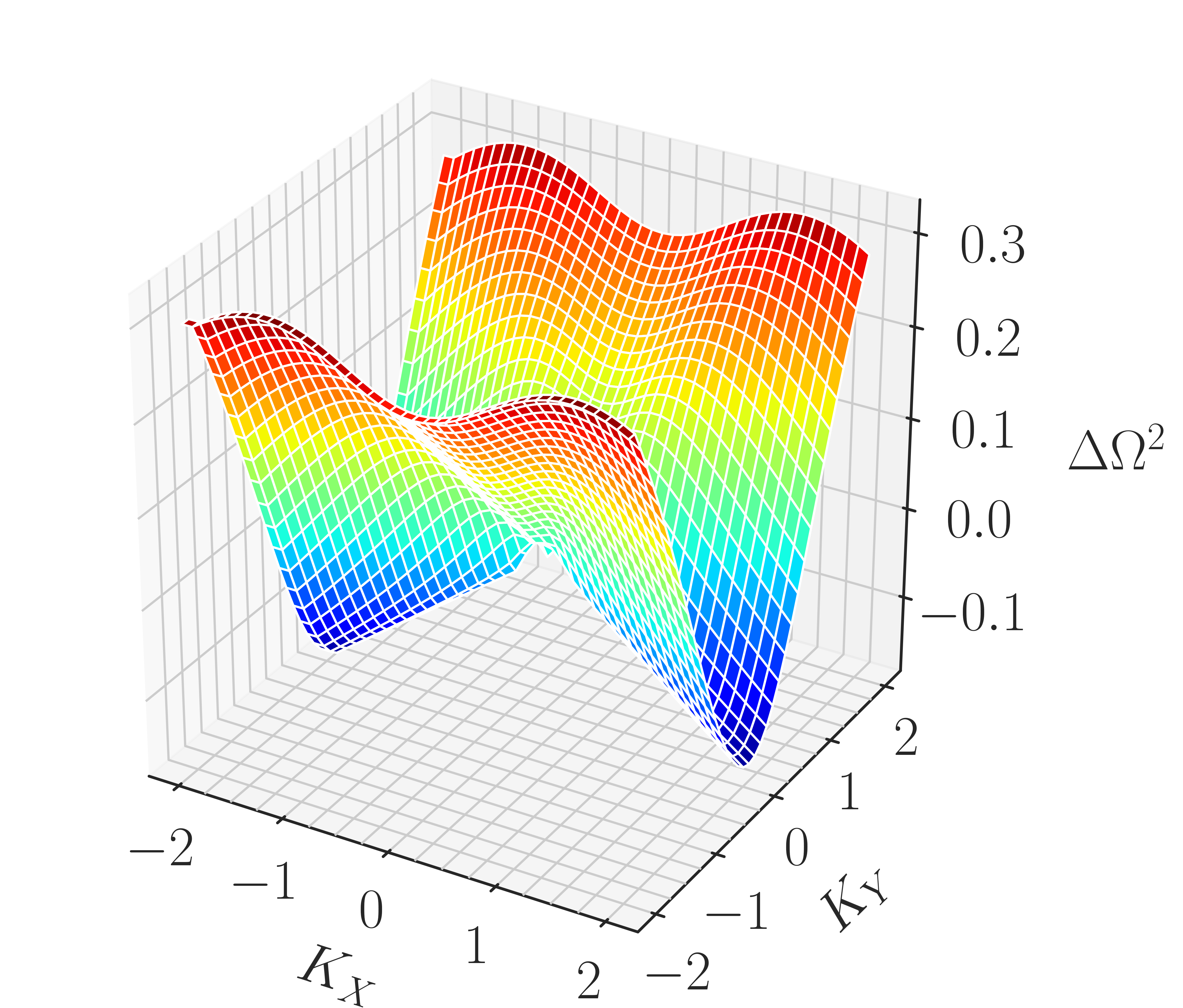}
    \caption{$\Delta\Omega^2 = \Omega^2 - \Omega_\mathrm{hydro}^2$ is plotted for a magnetized polytrope, with $\epsilon=0.1$, $m=20$, $n=2.5$, and $\gamma=5/3$. This surface plot shows the difference between the two surface plots in \fig{fig:f2}.}
    \label{fig:f3}
\end{figure}

We first validate that, in the absence of the magnetic field, our solver successfully reproduces two branches of Lamb's exact analytical
dispersion relation [Eq.~\eqref{eq:lambfull}]. 
We identify the pressure-driven modes by comparing their eigenfrequencies with those predicted by Lamb's solution. We then impose a very weak background magnetic field and systematically increase its strength, and track changes to the eigenfrequencies. A case of such magnetic modification of pressure-driven waves is shown 
in Fig.~\ref{fig:f2}. Although, in Fig.~\ref{fig:f2}, we display two surface plots of $\Omega^2(K_X, K_Y)$---one for 
the hydrodynamic and other for the magnetized polytrope---the two plots are visually indistinguishable. The 
difference between the two surface plots is shown in Fig.~\ref{fig:f3}.  For $K_Y=0$, $\Delta \Omega^2$ is 
negative, and for $K_X=0$, $\Delta \Omega^2$ is positive and relatively large. We also note a minor decrease in 
positive value of $\Delta \Omega^2$ in going from $K_X\approx 0$ to $K_X=0$.

\begin{figure}
    \centering
    \includegraphics[width=0.9\columnwidth]{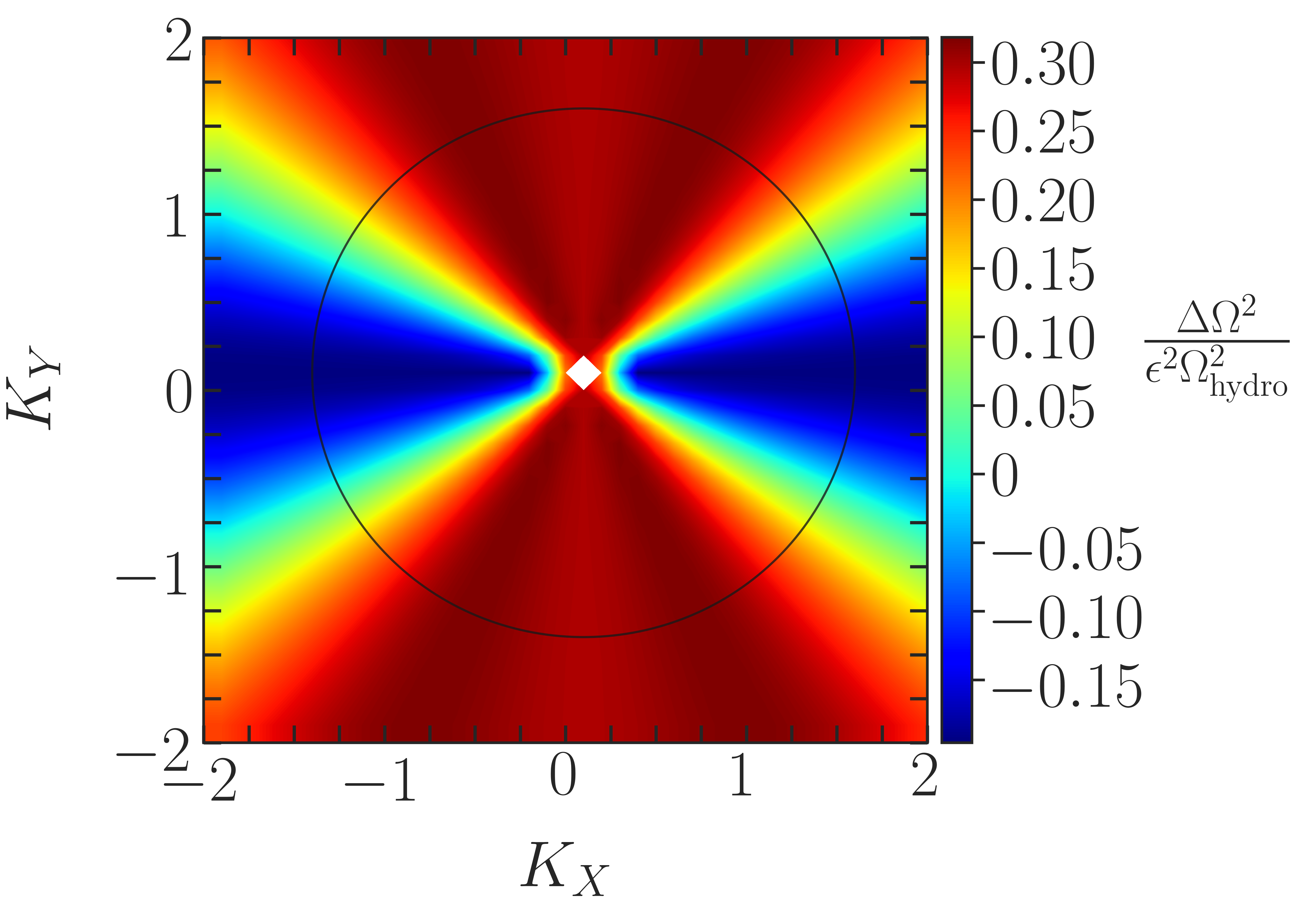}
    \caption{Relative difference of squared-frequency depends on the wavevector propagation angle (spanned by the black circle), but the relative difference is insensitive to the wavevector magnitude. The parameters chosen are $\epsilon=0.1$, $m=20$, $n=2.5$, and $\gamma=5/3$. }
    \label{fig:f4}
\end{figure}

It turns out that $\Delta \Omega^2$ is related to the hydrodynamic squared-frequency $\Omega_\mathrm{hydro}^2$; and $\Delta \Omega^2/\Omega_\mathrm{hydro}^2$ is almost entirely independent of the wavevector magnitude (Fig.~\ref{fig:f4}). Only angular dependence is observed.  

The extremely low wavenumbers in Fig.~\ref{fig:f4} correspond to very large scale waves that cannot be captured in a finite box in numerical calculation. To capture lower wavenumbers, we significantly extend the domain size along the vertical $z$-axis, which allow us to obtain fully converged numerical results for other wavenumbers.

\section{Reduction to wave equation}
\label{sec:nondim}
To analytically determine the dispersion relation, we solve the set of equations
\eqref{eq:linux2}--\eqref{eq:linuz2} perturbatively in the limit of a weak
magnetic field, i.e.,
$\epsn \ll 1$.
By setting $\epsn=0$, we  recover the Lamb's equations for the unmagnetized polytrope ~\citep{lamb1911}.
The Lamb's equations reduce to a second-order differential equation for $\hat{\chi}$.
With the same goal, we proceed in the following manner.
We rewrite Eqs.~\eqref{eq:linux2}--\eqref{eq:linuz2} as
\begin{equation} \label{eq:matform1}
\begin{bmatrix}
    \begin{array}{c c c}
        M_{11} & M_{12} & M_{13}  \\ 
        M_{21} & M_{22} & M_{23} \\ 
        M_{31} & M_{32} & M_{33} \\ 
    \end{array}
\end{bmatrix}
\begin{bmatrix}
    \begin{array}{c}
        \hat{u}_x  \\ 
        \hat{u}_y \\ 
        \hat{u}_z \\ 
    \end{array}
\end{bmatrix}
=
\begin{bmatrix}
    \begin{array}{c}
        h_x(\hat{\chi})  \\ 
        h_y(\hat{\chi}) \\ 
        h_z(\hat{\chi}, \partial_{Z}\hat{\chi}) \\ 
    \end{array}
\end{bmatrix},\\
\end{equation}
where the matrix elements $M_{ij}$ are independent of $Z$-derivatives, and are functions of
$K_X, K_Y, \Omega, n, \gamma, \epsn,$ and $Z$ only (for their complete expressions, see Appendix~A).
In arriving at Eq.~\eqref{eq:matform1}, we have replaced $\partial_Z \hat{u}_x$ in the first term on the
left-hand side of Eq.~\eqref{eq:linuz2} with its exact expression obtained by differentiating
Eq.~\eqref{eq:linux2} with respect to $Z$.
We then substitute $\partial_Z$ operation in $\partial_Z \hat{u}_z$ by writing it as
$\hat{\chi} - i K_X \hat{u}_x - i K_Y \hat{u}_y$. Such a process removes $\partial_Z$ operation from the matrix $M$.
The functions $h_{\nu}$ are linear in $\hchi$ and $\partial_Z\hchi$. 

Straightforward inversion of the matrix $M$ expresses all components of the velocity in terms of
$\hat{\chi}$ and $\partial_Z \hat{\chi}$:
\begin{equation}\label{eq:uvecchi}
    \hat{u}_{\nu} = f_{\nu}(Z) \hchi + g_{\nu}(Z) \partial_Z \hchi,
\end{equation}
where $\nu$ can be either $x,y,$ or $z$. We recognize that $f_{\nu}$ and $g_{\nu}$ are functions of $Z$,
but do not involve $\partial_Z$.
The three velocity components of Eq.~\eqref{eq:uvecchi} can now be subsumed into a single second-order
differential equation for $\hchi$:
\begin{equation}
  \label{eq:schrodinger1}
    \partial_Z^2 \hchi + P(Z, \epsn) \partial_Z \hchi + R(Z, \epsn)\hchi = 0,
\end{equation}
which can be recast into the normal form of the second-order differential equation by changing variable as 
\begin{equation}
	\hchi(Z) = \hat{\psi}(Z) \exp{\left[-\frac{1}{2}\int_{Z} dZ\, P(Z)\right]},    
\end{equation}
which reduces Eq.~\eqref{eq:schrodinger1} to the wave equation
\begin{equation}
  \label{eq:schrodinger2}
    \partial_Z^2 \hat{\psi} + \Gamma^2(Z, \epsn,\delta)\hat{\psi} = 0.
\end{equation}
In \eq{eq:schrodinger2}, we use the explicit notation
  $\Gamma^2(Z, \epsn,\delta)$ to remind us that the two small parameters $\epsn$
  and $\delta$ are implicit in $\Gamma$,
  where $\delta=K/\Omega^2$, with $K=\sqrt{K_X^2+K_Y^2}$ representing the
  magnitude of the wavevector and $\Omega$ representing the eigenfrequency.

The procedure outlined above appears straightforward.
However, the analytical manipulations in arriving at Eq.~\eqref{eq:schrodinger2}---a magnetized version of
Lamb's equation---require laborious and careful calculations, as $\Gamma^2(Z, \epsn,\delta)$ alone conceals an
expression of exhaustive length---tens of pages of this article.
In the absence of the magnetic field, the expression for
$\Gamma^2(Z, \epsn=0,\delta)=\Gn^2(Z,\delta)$
is beautifully short,
$\Gn(Z,\delta)= K^2(Z-\alpha)(\beta-Z)/Z^2$,
where $\alpha$ and $\beta$ are the two turning points---two zeros of $\Gn(Z,\delta)$---that depend on $\delta$, the property of the eigenfrequency. 

\section{Perturbative solution for anisotropic magnetic effect}
\label{sec:pertsol}

The presence of a weak magnetic field, characterized by small $\epsn$, may be considered as a perturbation to the
Lamb's two-turning-point eigenvalue problem.
Hence, the magnetic field changes both the locations of the turning points and
the form of the potential $\Gamma(Z, \epsn,\delta)$.

\subsection{Perturbative calculations}

A formal asymptotic solution to Eq.~\eqref{eq:schrodinger2} is
  constructed by introducing a book-keeping, small
  parameter~\footnote{The parameter $\zeta$ can replaced with unity at
    the outset or after performing the $\zeta$-expansion.} $\zeta$ in
  \eq{eq:schrodinger2}~\citep[see e.g.,][]{bender1978, tripathi2022b}:
\begin{equation}
  \label{eq:schrodinger_zeta}
    \zeta^2 \partial_Z^2 \hat{\psi} + \Gamma^2(Z, \epsn,\delta)\hat{\psi} = 0.
\end{equation}
To find the magnetically modified eigenfrequencies, we develop below an
asymptotic theory starting from Eq.~\eqref{eq:schrodinger_zeta}, by using
perturbative expansions in powers of $\zeta$, $\epsn$, and $\delta$,
in that order.


Following the standard procedure of the $\zeta$-expansion of the state
vector $\hat{\psi}$, we write the JWKB quantization
condition~\citep[see e.g.,][]{bender1978, tripathi2022b},
\begin{equation}
  \label{eq:zeta1}
    \frac{1}{\pi}\int_{Z_{1}(\epsn,\delta)}^{Z_{2}(\epsn,\delta)}\Gamma(Z, \epsn,\delta)\,dZ \sim \left(m+\frac{1}{2}\right);\,\,\,m=0,1,2,...,
\end{equation}
where $m$ refers to the eigenstate index, $\Gamma(Z, \epsn,\delta)$ is the magnetically modified wavenumber,
and $Z_1(\epsn,\delta)$ and $Z_2(\epsn,\delta)$ are the magnetically shifted
turning points.
Equation~\eqref{eq:zeta1} is the
leading-order contribution in the conventional JWKB asymptotic series,
  where the terms corresponding to the first two orders in $\zeta$ are
  retained.
  The first-order term yields $1/2$, shown on the right.
  Higher-order terms in $\zeta$ can be computed, see, e.g., Eqs.~(7) to (10)
  of \cite{tripathi2022b}.


Next, we expand the wavenumber $\Gamma(Z, \epsn,\delta)$ as
\begin{equation}
  \label{eq:Gammaexpansion}
\begin{aligned}
  \Gamma(Z, \epsn,\delta) = \Gn(Z,\delta) + \epsn \Gamma_1(Z,\delta)  &+ \epsn^2 \Gamma_2(Z,\delta)  \\
  &+ \epsn^3 \Gamma_3(Z,\delta)
  + \mathcal{O}{(\epsn^4)}.
\end{aligned} 
\end{equation}
Note that the leading-order effect of the Lorentz force on the wavefrequency appears only at the second
order $\mathcal{O}{(\epsn^2)}$ in the expansion, i.e.,
$\Gamma_1(Z,\delta)$ above is zero.
The expressions for $\Gamma_0(Z,\delta)$ and $\Gamma_2(Z,\delta)$ are
\begin{subequations}
\begin{align}
    \Gamma_0(Z,\delta) &= \frac{K \sqrt{(Z-\alpha)(\beta-Z)}}{Z},\\
    \Gamma_2(Z,\delta) &= \frac{(b_2 Z^2 + b_1 Z + b_0)}{\sqrt{(Z-\alpha)(\beta-Z)}}
    \label{eq:Gamma2_expression_defined},
\end{align}
\end{subequations}
where we have used $K_X = K \cos\theta$ and $K_Y=K\sin\theta$.
The parameters $\alpha$ and $\beta$ depend on $\delta=K/\Omega^2$ and satisfy the following properties
\begin{subequations}
\begin{align} \label{eq:alphabeta}
    \alpha \beta &= \frac{n(n+2)}{4 K^2},\\
    \alpha + \beta &=  \left[\frac{\Omega^2}{K} -
      \frac{(n+1)(n+1-n\gamma)}{\gamma^2} \frac{K}{\Omega^2}\right]\frac{1}{K}.
\end{align}
\end{subequations}
The lengthy expressions of $b_0$, $b_1$, and $b_2$ are presented in Appendix~B. We note that these parameters depend on $\Omega, K, \theta, n,$ and $\gamma$ only.

Now we expand the turning points $Z_1(\epsn,\delta)$ and $Z_2(\epsn,\delta)$
around the turning points of the unmagnetized polytrope,
$Z_1^{(0)} = \alpha$ and $Z_2^{(0)} = \beta$ in powers of $\epsn$.
For clarity we now suppress the implicit $\delta$--dependence of the
  turning points.
For example, we write $Z_j(\epsn,\delta)$ as $Z_j(\epsn)$, which
we expand in powers of $\epsn$ as
\begin{equation} \label{eq:TPexpansion}
    Z_j(\epsn) = Z_j^{(0)} + \epsn Z_j^{(1)} + \epsn^2 Z_j^{(2)}  + \mathcal{O}{(\epsn^3)},
\end{equation}
where $j=1$ and $j=2$ refer to the left and the right turning points, respectively (i.e., $\alpha < \beta$).
We note that, in Eq.~\eqref{eq:TPexpansion}, the correction term at the first order in $\epsn$ is zero, i.e.,
$Z_j^{(1)}=0$.
We find this result by substituting the expression for $Z_j(\epsn)$ from Eq.~\eqref{eq:TPexpansion} in
$\Gamma(Z, \epsn) = 0$, and by employing Eq.~\eqref{eq:Gammaexpansion}.
Solving the resulting equation order-by-order in $\epsn$ produces
$Z_j^{(1)}=0$.
We note, however, that, at the second order in $\epsn$ (which is where the effect of the Lorentz force
comes in action), $Z_j^{(2)}$ becomes non-zero. The expressions for $Z_j^{(2)}$ are given in Appendix~C. 

Because the term $Z_j^{(2)}$ appears at the second order in $\epsn$, it may be tempting to assume that the
term contributes to second order in $\epsn$ itself in the JWKB integral in
\eq{eq:zeta1}.
This, however, is not the case.
The term contributes to third and higher orders in $\epsn$ as we show next. 
 Expanding Eq.~\eqref{eq:zeta1},
 \begin{equation}
   \label{eq:quant2}
\begin{aligned}[b]
    \frac{1}{\pi}\int_{Z_{1}^{(0)} + \epsn^2 Z_1^{(2)} + \mathcal{O}{(\epsn^3)} }^{Z_{2}^{(0)} + \epsn^2 Z_2^{(2)} + \mathcal{O}{(\epsn^3)} }&\left[\Gamma_0(Z)+\epsn^2 \Gamma_2(Z)+\mathcal{O}{(\epsn^3)}\right]\,dZ\\
    &\sim \left(m+\frac{1}{2}\right);\,\,\,m=0,1,2,... \/.
\end{aligned}
\end{equation}
We now integrate $\Gn(Z)$ as
\begin{equation}
  \label{eq:quant3}
\begin{aligned}[b]
    &\int_{Z_{1}^{(0)} + \epsn^2 Z_1^{(2)} + \mathcal{O}{(\epsn^3)} }^{Z_{2}^{(0)} + \epsn^2 Z_2^{(2)} + \mathcal{O}{(\epsn^3)} }\Gn(Z)\,dZ \\
    &= \left( \int_{\alpha + \epsn^2 Z_1^{(2)} + \mathcal{O}{(\epsn^3)} }^{\alpha } + \int_{\alpha }^{\beta} + \int_{\beta}^{\beta+ \epsn^2 Z_2^{(2)} + \mathcal{O}{(\epsn^3)} }\right)\Gn(Z)\,dZ\\
    &=-\frac{2\epsn^3 K (\beta-\alpha)^{1/2} \left[Z_1^{(2)}\right]^{3/2}}{3\alpha} + \int_\alpha^\beta\Gn(Z)\,dZ \\
    &\hspace{2cm}+ \frac{2\epsn^3 K (\beta-\alpha)^{1/2} \left[-Z_2^{(2)}\right]^{3/2}}{3\beta} + \mathcal{O}{(\epsn^4)}\/,
\end{aligned}
\end{equation}
where we notice terms with $\epsn^3$ arising from the second-order shifts in the turning points, $\epsn^2 Z_1^{(2)}$ and $\epsn^2 Z_2^{(2)}$.
The additional power of $\epsn$ emerges from the integrand
$\Gn(Z)$, which has a term $\sqrt{Z-Z_j^{(0)}}$. This term when expanded around $Z_j^{(0)}$ in powers of $\epsn$ contributes an $\epsn$ to the integral. 

In Eq.~\eq{eq:quant3}, the term $\int_\alpha^\beta\Gamma_0(Z)\,dZ$ is the integral 
that one finds in Lamb's calculations. Since here we seek the effect of magnetic fields on Lamb waves, we replace $\int_\alpha^\beta\Gamma_0(Z)\,dZ$  with Lamb's exact dispersion relation for hydrodynamic waves [Eq.~\eqref{eq:lambfull}]:
\begin{equation}
  \label{eq:lambterm}
  \begin{aligned}[b]
	  & \frac{1}{\pi}\int_\alpha^\beta\Gamma_0(Z)\,dZ \equiv \\
	  &I_\mathrm{Lamb} = \frac{(n+1)}{2} \left[\frac{\Omega^2}{K(n+1)} + \frac{(\gamma n - n - 1)K}{\gamma^2 \Omega^2} - 1 \right] + 1.
  \end{aligned}
  \end{equation}
Thus, we replace the integral $\int_\alpha^\beta \Gamma_0(Z)\,dZ $, appearing in Eq.~\eqref{eq:quant3}, with
$I_\mathrm{Lamb}$ from Eq.~\eqref{eq:lambterm} to obtain
\begin{equation}
  \label{eq:quant4}
\begin{aligned}
    I_\mathrm{Lamb} &+ \frac{\epsn^2 }{\pi} \int_\alpha^\beta \Gamma_2(Z)dZ+\mathcal{O}{(\epsn^3)}\\
    &\hspace{1cm}\sim \left(m+\frac{1}{2}\right);\,\,\,m=0,1,2,...,
\end{aligned}
\end{equation}
which is accurate up to second order in $\epsn$.
We substitute $\Gamma_2(Z)$ from Eq.~\eqref{eq:Gamma2_expression_defined} and
perform the integral in Eq.~\eq{eq:quant4} to arrive at
\begin{equation}\label{eq:quant_for_numerics}
\begin{aligned}[b]
I_\mathrm{Lamb} - \epsn^2 &\left[b_0 + b_1 \left(\frac{\alpha+\beta}{2}\right) + b_2 \left( \frac{3 \alpha^2 + 2\alpha\beta+ 3 \beta^2}{8} \right) \right]\\
&\hspace{1cm}\sim \left(m+\frac{1}{2}\right);\,\,\,m=0,1,2,...\,.
\end{aligned}
\end{equation}


Employing fast-wave approximation, we now expand each term on the left-hand
side of Eq.~\eqref{eq:quant_for_numerics} in powers of $\delta = K/\Omega^2$ as
\begin{subequations}
\begin{align}
    I_\mathrm{Lamb} &\approx \frac{1}{\delta}\left[\frac{1}{2} + \frac{\delta(1-n)}{2} + \mathcal{O}(\delta^2) \right],\\
    b_0 &\approx \frac{1}{\delta}\left[\frac{2-\gamma}{4} + \mathcal{O}(\delta^2) \right]\label{eq:b0approx},\\
    b_1 \left(\frac{\alpha+\beta}{2}\right) &\approx \frac{1}{\delta}\left[\frac{\cos^2\theta}{4} + \mathcal{O}(\delta^2) \right], \label{eq:b1approx}\\
    b_2 \left( \frac{3 \alpha^2 + 2\alpha\beta+ 3 \beta^2}{8} \right) & \approx  \frac{1}{\delta}\left[\frac{-3\cos^4\theta}{8} + \mathcal{O}(\delta^2) \right] \label{eq:b2approx}.
\end{align}
\end{subequations}

Thus we obtain a simplified dispersion relation from Eq.~\eqref{eq:quant_for_numerics}:
\begin{equation} \label{eq:quant_notexcellent}
\begin{aligned}[b]
    &\frac{\Omega^2}{K} \left[\frac{1}{2} - \frac{\epsn^2}{4} \left\{2-\gamma+\cos^2\theta -\frac{3\cos^4\theta}{2} \right\} \right] \\
    &\hspace{2cm}\sim \left(m+\frac{n}{2}\right);\,\,\,m=0,1,2,...\,.
\end{aligned}
\end{equation}
It turns out that Eq.~\eq{eq:quant_notexcellent} does not agree excellently with our
numerical results, see Fig.~\ref{fig:f6} in Appendix~D.
But replacing $\cos^2\theta$ with $(\cos^2\theta)/2$ gives
excellent agreement. 

Informed in this way, we write the final dispersion relation
\begin{equation}
  \label{eq:quantfinal}
\boxed{
\begin{aligned}[b]
    &\frac{\Omega^2}{K} \left[\frac{1}{2} - \frac{\epsn^2}{4} \left\{2-\gamma+\frac{\cos^2\theta}{2}-\frac{3\cos^4\theta}{2} \right\} \right] \\
    &\hspace{2cm}\sim \left(m+\frac{n}{2}\right);\,\,\,m=0,1,2,...\,.
\end{aligned}
}
\end{equation}

\subsection{Comparison between theory and numerics}
\label{sec:comp}
\begin{figure*}
    \centering
    \includegraphics[width=1.2\columnwidth]{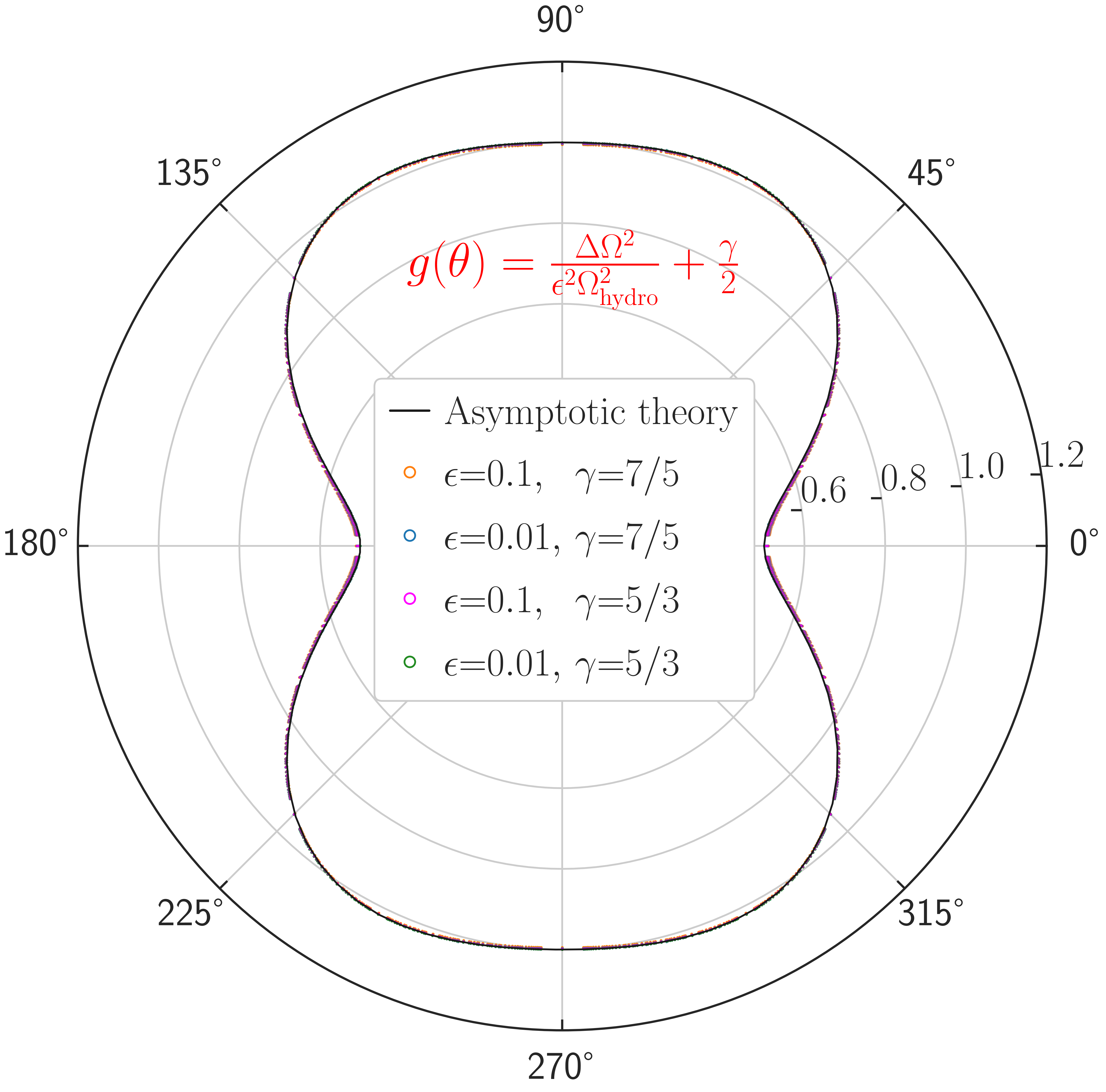}
    \caption{\textbf{Theory and numerics:} Our theory predicts that
      $g(\theta) = \Delta \Omega^2/(\epsn^2\Ohydro^2) + \gamma/2$ depends only
      on $\theta$ as given in Eq.~\eq{eq:gtheta}.
      The function $g(\theta)$ is independent of the wavenumber $K$, Alfv\'enic Mach number $\epsilon^{-1}$, the polytropic index $n$,
      the adiabatic index $\gamma$, and the eigenstate index $m$.
      Here, we plot $g(\theta)$ obtained from our numerical
      solutions for several different values of $m$, ranging from $10$ to $35$, and for several different values of
      $K$ ($0.6\leq |K| \leq 2$), and for two values of $\epsn$ and two values of $\gamma$ 
      (different symbols).
      All data points collapse on the same universal curve.
      The function $g(\theta)$ from our asymptotic theory, Eq.~\eq{eq:gtheta},
      is also plotted here. The asymptotic curve is indistinguishable from the exact
      numerical solutions. }
    \label{fig:f5}
\end{figure*}

Using the Lamb's relation, $\Ohydro^2 \sim K(2m+n)$, we rewrite
Eq.~\eq{eq:quantfinal} as
\begin{equation}\label{eq:gtheta}
\begin{aligned}[b]
  g(\theta) &\equiv \gamma/2+\Delta \Omega^2/(\epsn^2\Ohydro^2) \\
            &= 1 + \frac{1}{4}\left(\cos^2\theta - 3\cos^4\theta \right),
\end{aligned}
\end{equation}
where $\Delta \Omega^2 \equiv \Omega^2 - \Ohydro^2$.
This is a remarkable result.
The right-hand-side of Eq.~\eq{eq:gtheta} is independent of every other
possible parameter other than $\theta$ !
In \fig{fig:f5}, we plot the function $g(\theta)$ from our numerical
determination of the eigenfrequencies for numerous different
values of $m$ and $K$, and two different values of $\epsn$, and two different values of $\gamma$.
These numerical values are shown with different markers.
All the different curves collapse onto each other, creating a universal master curve.
We also plot our asymptotic expression, Eq.~\eqref{eq:quantfinal}, which agrees very well with the
numerical results.
This demonstrates that our theory is in excellent agreement with
numerics even for $\epsn$ as large as $\epsn = 0.1$. 

It is surprising that, in our attempt to capture the anisotropy
brought in by the magnetic field, despite a myriad of unwieldy expressions
encountered on the way, an expression as simple as Eq.~\eqref{eq:quantfinal},
is obtained.
This simple expression is also highly accurate, as demonstrated in \fig{fig:f5}.
The gratifying success of our asymptotic theory is somewhat unexpected,
given that the analytical solution is accurate only up to the leading order in
$\zeta, \epsn,$ and $\delta=K/\Omega^2$.  

More accurate solutions may be obtained by using the full expressions of
$b_0, b_1,$ and $b_2$ from Appendix~B in
Eqs.~\eqref{eq:b2approx}--\eqref{eq:b0approx}, and retaining higher-order terms
in $\zeta$ and $\epsn$.

We emphasize that the closed-form expression we obtain
is not solely by use of the perturbative calculation.
At the last step, we needed guidance from our numerical solutions.
This implies that future work should develop a better perturbation theory.
  We discuss a few possibilities. 
  Let us first revisit our perturbative calculation.
  Equation~\eqref{eq:schrodinger2} is exact.
  It has two small parameters, $\epsn$ and $\delta$.
  For $\epsn=0$, this equation reduces to the Lamb's problem.
  We note that our final
  dispersion relation \eq{eq:quantfinal} reproduces the high-frequency Lamb's dispersion relation~\eqref{eq:lambasymp} 
  if we set $\epsn=0$ in Eq.~\eqref{eq:quantfinal}.
  In our perturbative calculation, we also introduce
  the additional small parameter $\zeta$
  to allow us to use the JWKB formalism.
  The first step then is to retain the first two leading-order terms in $\zeta$.
  Thus we obtain Eq.~\eqref{eq:zeta1}. 
  We arrive at a curious JWKB problem where the turning points themselves
  depend on a small parameter $\epsn$, see Eq.~\eqref{eq:quant2}.
  We retain terms up to second order in $\epsn$ in this expansion,
  thereby arriving at Eq.~\eq{eq:quant4}.
  Next, we expand all terms in powers of $\delta$.
  If we change the order of expansion in $\epsn$ and $\delta$, then
  we no longer obtain the mathematical structure of the
  two-turning-point JWKB problem.
  The leading-order term reduces to an expression whose $Z$-dependence is
  solely $Z^{-1/2}$.
  To reformulate that expansion such that we can compare it with
  the Lamb's result, we must resum the $\delta$-expanded infinite series
  which is not straightforward.
  So, this does not seem to be a useful alternative. 
    Next, we remind the reader that a naive application of the JWKB method to
  calculate
  eigenvalues of the Schrodinger equation for either the harmonic
  oscillator or the Coulomb potential in three dimensions gives a factor
  of $\ell(\ell+1)$ [$\ell$ is the quantum number for orbital angular momentum],
  which must be replaced with $(\ell+1/2)^2$ to obtain the
  correct expression~\citep{langer1937connection}.
  This correction appears naturally when the JWKB series is resummed to all
  orders~\citep{romanovski2000, robnik1997,tripathi2022b}.
  Hence, there is a possibility that a resummation of  the
  higher-order terms in powers of $\zeta$ may reproduce
  Eq.~\eqref{eq:quantfinal}.
  Finally, we note the third possibility.
  We mapped Eq.~\eqref{eq:schrodinger1} to a form amenable to the JWKB
  method.
  Instead, for $\epsn=0$, Eq.~\eqref{eq:schrodinger1} can be transformed to the
  confluent hypergeometric equation---this is how Lamb first solved for the waves in a hydrodynamic polytrope. 
  Thus, for a small $\epsn$, it possible to use the method of multiple
  scales in a way such that the confluent hypergeometric functions
  appear as leading-order solutions, allowing to compute perturbed eigenfrequencies.

\subsection{Role in helioseismic inversions}
Since our theoretical formula and numerics show in Fig.~\ref{fig:f5} that
  the expression
  $\Delta \Omega^2/(\epsn^2\Ohydro^2)+\gamma/2$
  collapses on a single universal curve
  $g(\theta)=1 + \frac{1}{4}\left(\cos^2\theta - 3\cos^4\theta \right)$,
  it suggests that high-frequency helioseismic oscillation data
  should collapse on a single curve if one plots the fractional change in observed
  frequency compared to the Lamb's hydrodynamic model.
  Such a plot allows extraction of two parameters:
  $\epsn^2$, which encodes the magnetic field strength, and $\gamma$, which is
  the adiabatic index of the gaseous atmosphere.
  Since our theory is applicable to high-frequency ($m\gg 1$) waves, the
  polytropic index $n$ cannot be inferred reliably using our method,
  as the squared-frequency of oscillations in the hydrodynamic model in
  Eq.~\eqref{eq:lambasymp} scales as $2m+n \sim 2m$.

We also note that we have considered a horizontal magnetic field
  in local Cartesian coordinates, which corresponds to non-radial magnetic fields
  in spherical coordinates.
  Such non-radial fields are found to be the most sensitive to the helioseismic
  inversion kernels near the solar surface~\citep{das2022recipe,
    das2020sensitivity}. This justifies our consideration.
  Note further that the inversion kernels in general are model-dependent,
  and the models for stars other than the Sun are not well-constrained.
  Since our formula has only two fitting parameters, more reliable estimation of those parameters is possible.

The effect of magnetic fields on slow gravity-driven waves is not
    considered in this paper, as the gravity-driven waves generally penetrate
    deep in the core of a star, which requires global geometry.
    Our current analysis in the Cartesian domain is appropriate for the
    fast pressure-driven waves that are confined in the subsurface layers
    of a star.
    We shall nevertheless remark briefly on the potential impact of magnetic
    fields on slow gravity-driven waves.
For the slow waves, the first term on the left-hand side of
Eq.~\eqref{eq:lambfull} may be neglected, which implies:
(i) squared time period of such waves is linearly proportional to the
eigenstate index, and
(ii) such waves in unmagnetized medium become
unstable when $\gamma < (1+1/n)$.
This instability criterion (ii) is consistent with the energy principle of \cite{newcomb1961}.
When a magnetic field is present, by applying the  energy principle
of \citeauthor{newcomb1961}, we find, for
$K_X\neq 0$,\footnote{For $K_X=0$, the criterion for the gravity waves to
  become unstable is slightly modified: $\gamma < (1+1/n)(1+1/\beta)-2/\beta$.}
the instability threshold on $\gamma$ for the gravity waves is lifted to
$\gamma < (1+1/n)(1+1/\beta)$.
When the magnetic field is very strong ($\beta\ll 1$), the regular perturbation
series in powers of $\epsn\, (\propto 1/\sqrt{\beta})$ is possibly inadequate
for unstable gravity-driven waves.
Such considerations are clearly beyond the scope of the present paper.
We note, however, that thorough understanding of the effect of magnetic
  fields on mixed gravito-acoustic waves holds a promise to deliver reliable
  estimation of the strengths and geometries of magnetic fields that are
  buried in deep layers of stellar interiors~\citep{bugnet2021,mathis2021,
    mathis2023}.
  As a testament to this promise, recent global asteroseismic studies have
  detected and characterized magnetic fields in the core of red giant stars,
  opening a wholly new avenue to magnetoseismology \citep{li2022, li2023,
    deheuvels2023}.

\section{Conclusions}\label{sec:conclusion}
Here we derive, for pressure-driven waves, an accurate and simple analytical
formula that captures the effects of magnetic field and five other parameters:
adiabatic index, polytropic index, eigenmode state index, wavenumbers, and the
angle between the wavevector and the magnetic field
[$\theta\mathrm{=}\cos^{-1}(\hat{\mathbf{K}}\cdot \hat{\BB}_0)$].
Such a six-parameter-dependent formula is distilled using a 
perturbative solution to magnetized version of Lamb's hydrodynamic polytropic
waves.
Our explicit analytical formula overcomes the limitation of previously
attempted formulae for the magnetized polytrope that were presented in general
integral forms; such a formulation, for instance, that of
\citeauthor{gough1990} [\citeyear{gough1990}; e.g., Eq.~(4.11)] and
\cite{bogdan1997}, requires numerical evaluation of the eigenfrequencies, and
thus leaves out the critical step of obtaining an analytical understanding and
expression.
We arrive at our result, guided by
our numerical solutions and at the cost of extensive use of Mathematica
for our perturbative analyses. 

The simplicity and accuracy of our formula are encouraging to employ the
formula to help solve the inverse problem of magnetoseismology.
Our formula provides an explicit, analytical dependence of the observed surface
oscillation frequency with the orientation and strength of the 
subsurface magnetic field.
Such an understanding may be able to predict the surface-emergence
  location and strength of active regions, and to understand the properties of
  subsurface dynamo, e.g., magnetic activity  in the near-surface shear
  layer \citep{vasil2024}.

Nonlinear asteroseismology can also directly benefit from our analytical work,
as weak turbulence theory of asteroseismic waves inevitably requires accurate
and simple expressions of linear wave frequencies in resonant triad
interactions.
The effect of the magnetic field on such waves remains unknown.
However, observations now exist that suggest resonant mode interactions are
possible and can be a critical element of strongly pulsating
stars \citep{guo2020}.
Nonlinear mode coupling of linear eigenmodes \citep{tripathi2023a,
  tripathi2023b,tripathi2024a} may also need to be analyzed, in addition to
mode resonances.
Future planned research will directly take advantage of the formula derived
here to assess the role of the magnetic field and other parameters in
asteroseismic wave turbulence, which is eagerly awaiting to soon enter adulthood
from its infancy.

\section*{Acknowledgments}
We are pleased to thank Ellen G.~Zweibel for helpful discussions and for her contribution in applying the energy principle of Newcomb (1961) to our system. We are grateful to Srijan B.~Das and the anonymous reviewer for their feedback that improved the paper. The inception of this paper took place at Nordic Institute for Theoretical Physics (NORDITA), Sweden, while B.T. was visiting on a research fellowship generously made available by NORDITA. 

\section*{Data Availability}
Mathematica notebook and Dedalus script used in this paper are deposited in\\
 \href{https://github.com/BindeshTripathi/polytrope}{https://github.com/BindeshTripathi/polytrope}. The notebook is developed to perform lengthy analytical manipulations, and the python script is created for spectral numerical solution of linear MHD waves. 

\clearpage

\onecolumngrid

\setcounter{equation}{0}
\setcounter{figure}{0}
\setcounter{table}{0}
\makeatletter
\renewcommand{\theequation}{A\arabic{equation}}
\renewcommand{\thefigure}{A\arabic{figure}}

\section*{Appendix~A}

The matrix elements $M_{ij}$ and $h_\mu$ that appear in Eq.~\eqref{eq:matform1} are
\begin{subequations}
\begin{align}
    M_{11} &= - \Omega^2(1+\gamma \epsn^2/2),\\
    M_{12} &= 0,\\
    M_{13} &=\frac{-i K_X (n+1) (1+\gamma \epsn^2/2)}{\gamma},\\
    M_{21} & = \epsn^2 K_X K_Y Z,\\
    M_{22} &= -\Omega^2(1+\gamma \epsn^2/2) + \epsn^2 K_X^2 Z,\\
    M_{23} &= \frac{-i K_Y (n+1) (1+\gamma \epsn^2/2)}{\gamma},\\
    M_{31} &=\frac{i K_X}{\gamma} \left[1 + \epsn^2\left( \frac{3\gamma }{2} + \frac{K_X^2 Z}{\Omega^2}\right) \right],\\
    M_{32} &= \frac{i K_Y}{\gamma} \left[1 + \epsn^2\left( \frac{\gamma }{2} + \frac{K_X^2 Z}{\Omega^2}\right) \right],\\
    M_{33} &= \frac{-\Omega^2(1+\gamma \epsn^2/2) + \epsn^2 K_X^2 Z}{n+1},\\
    h_x &= i K_X Z \hat{\chi},\\
    h_y &= i K_Y Z \hat{\chi} (1+\epsn^2),\\
    h_z &= \hat{\chi} + \frac{Z \partial_Z \hat{\chi}}{n+1} + \epsn^2 \hat{\chi} \left[1 + \frac{K_X^2 Z \left\{\gamma - (n+1)(1+\gamma \epsn^2/2) \right\}}{(n+1)\gamma\Omega^2(1+\gamma \epsn^2/2)} \right] + \epsn^2 \frac{Z \partial_Z \hat{\chi}}{n+1} \left[1+\frac{K_X^2 Z}{\Omega^2(1+\gamma \epsn^2/2)} \right].
\end{align}
\end{subequations}

\section*{Appendix~B}

The parameters introduced in Eq.~\eqref{eq:Gamma2_expression_defined}, while writing the expression for $\Gamma_2(Z)$, appear below:
\begin{widetext}
\begin{subequations}
\begin{align}
    b_2 &= \frac{-K^3 \cos^4\theta}{\Omega^2},\\
    b_1 &= \frac{-K \left[-\Omega^8\gamma^4 +  \Omega^4\gamma^2 K^2 (n+1)\left\{(n+1+\gamma-\gamma n)\cos(2\theta) - n-1-3\gamma\right\} + 2K^4 (n+1)^4 - K^4 (n+1)^3(n+1-\gamma n)\cos(2\theta)\right]}{2\Omega^4\gamma^2 \mathrm{sec}^{2}\theta\, \left[\Omega^4\gamma^2- K^2(n+1)^2 \right]},\\
    b_0 &= \frac{\Bigg(\begin{gathered}4\Omega^8\gamma^4 (2-\gamma) - 2 \Omega^4 \gamma^2 K^2 \left\{2(n+1)^2 (4-3\gamma) + n(2n+3)\gamma^2 \right\} + K^4 (n+1)^2 \left\{n\gamma^2 (2n+3) -8(n+1)^2(\gamma-1) \right\}\\
    + 2K^2 \gamma \cos(2\theta)\left[K^2(n+1)^2\left\{n(\gamma-4+2n(\gamma-1)) -2 \right\} + \Omega^4 \gamma^2 \left\{2(n+1)^2 -\gamma n(2n+3)\right\} \right] + n\gamma^2 K^4 (n+1)^2 \cos(4\theta)  \end{gathered}\Bigg)}{16\Omega^2\gamma^2 K \left[\Omega^4\gamma^2- K^2(n+1)^2 \right]}
\end{align}
\end{subequations}
\end{widetext}

\section*{Appendix~C}
Due to the magnetic field, the locations of the turning points, $Z_1$ and $Z_2$, shift---which to the second order in $\epsn$ in Eq.~\eqref{eq:TPexpansion} are given by $Z_1^{(2)}$ and $Z_2^{(2)}$:
\begin{subequations}
\begin{align}
    Z_1^{(2)} &= \frac{\alpha (b_2 \alpha^2 + b_1 \alpha + b_0) }{K(\beta-\alpha)},\\
    Z_2^{(2)} &= \frac{-\beta (b_2 \beta^2 + b_1 \beta + b_0) }{K(\beta-\alpha)}.
\end{align}
\end{subequations}

\section*{Appendix~D}
\begin{figure}
    \centering
    \includegraphics[width=1\columnwidth]{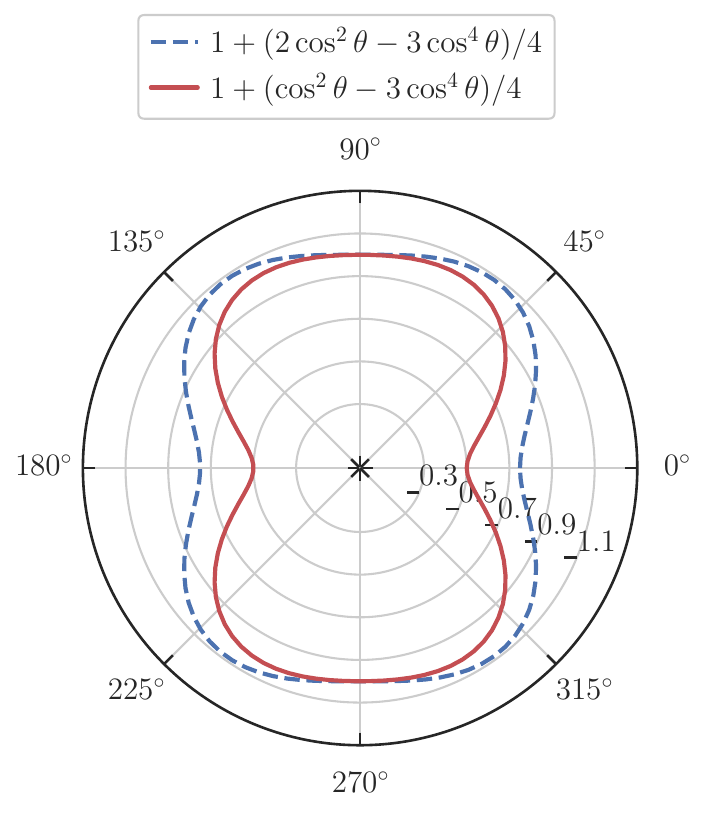}
    \caption{Comparison of different expressions to assess their individual contribution. The innermost (red) curve is the universal curve on which all numerical data collapse.}
    \label{fig:f6}
\end{figure}

\bibliography{references_aastex}{}
\bibliographystyle{aasjournal}

\end{document}